\documentclass[epj]{svjour}
\usepackage{multirow,rotating}
\usepackage{graphicx}

\newcommand{\atomki}{1}
\newcommand{\debrecen}{2}
\newcommand{\dresden}{3}
\newcommand{\infnpd}{4}
\newcommand{\unipd}{5}
\newcommand{\infnge}{6}
\newcommand{\lngs}{7}
\newcommand{\infnto}{8}
\newcommand{\infnmi}{9}
\newcommand{\unina}{10}
\newcommand{\infnna}{11}
\newcommand{\bochum}{12}
\newcommand{\teramo}{13}
\newcommand{\caserta}{14}

\begin{document}

\title{An actively vetoed Clover $\gamma$-detector for nuclear astrophysics at LUNA}
\author{
	T.\,Sz\"ucs\inst{\atomki,\debrecen} 
	D.\,Bemmerer \inst{\dresden}\thanks{e-mail: d.bemmerer@fzd.de} \and
	C.\,Broggini \inst{\infnpd} \and
	A.\,Caciolli \inst{\infnpd,\unipd} \and
	F.\,Confortola \inst{\infnge} \and
	P.\,Corvisiero \inst{\infnge} \and
	Z.\,Elekes \inst{\atomki} \and
	A.\,Formicola \inst{\lngs} \and
	Zs.\,F\"ul\"op \inst{\atomki} \and
	G.\,Gervino \inst{\infnto} \and
	A.\,Guglielmetti \inst{\infnmi} \and
	C.\,Gustavino \inst{\lngs} \and
	Gy.\,Gy\"urky \inst{\atomki} \and
	G.\,Imbriani \inst{\unina,\infnna} \and
	M.\,Junker \inst{\lngs} \and
	A.\,Lemut \inst{\infnge}\thanks{Present address: Lawrence Berkeley National Laboratory, Berkeley, USA} \and
	M.\,Marta \inst{\dresden} \and
	C.\,Mazzocchi \inst{\infnmi} \and
	R.\,Menegazzo \inst{\infnpd} \and
	P.\,Prati \inst{\infnge} \and
	V.\,Roca \inst{\unina,\infnna} \and
	C.\,Rolfs \inst{\bochum} \and
	C.\,Rossi Alvarez \inst{\infnpd} \and
	E.\,Somorjai \inst{\atomki} \and
	O.\,Straniero \inst{\infnna,\teramo} \and
	F.\,Strieder \inst{\bochum} \and
	F.\,Terrasi \inst{\infnna,\caserta} \and
	H.P.\,Trautvetter \inst{\bochum} \\
	(LUNA collaboration)
	}                     
\institute{
	Institute of Nuclear Research (ATOMKI), Debrecen, Hungary 
	\and
	University of Debrecen, Debrecen, Hungary 
	\and
	Forschungszentrum Dresden-Rossendorf (FZD), Dresden, Germany 
	\and
	INFN Sezione di Padova, Padova, Italy 
	\and 
	Dipartimento di Fisica, Universit\`a di Padova, Padova, Italy 
	\and
	Dipartimento di Fisica, Universit\`a di Genova, and INFN Sezione di Genova, Genova, Italy 
	\and
	INFN, Laboratori Nazionali del Gran Sasso, Assergi, Italy 
	\and 
	Dipartimento di Fisica Sperimentale, Universit\`a di Torino, and INFN Sezione di Torino, Torino, Italy 
	\and 
	Universit\`a degli Studi di Milano, and INFN Sezione di Milano, Milano, Italy 
	\and
	Dipartimento di Scienze Fisiche, Universit\`a di Napoli "Federico II", and INFN Sezione di Napoli, Napoli, Italy 
	\and
	INFN Sezione di Napoli, Napoli, Italy 
	\and
	Institut f\"ur Experimentalphysik III, Ruhr-Universit\"at Bochum, Bochum, Germany 
	\and
	Osservatorio Astronomico di Collurania, Teramo, Italy 
	\and
	Seconda Universit\`a di Napoli, Caserta, Italy 
}
%
\date{Version accepted by Eur.Phys.J. A, \today}

\abstract{
An escape-suppressed, composite high-purity germanium detector of the Clover type has been installed at the Laboratory for Underground Nuclear Astrophysics (LUNA) facility, deep underground in the Gran Sasso Laboratory, Italy. The laboratory $\gamma$-ray background of the Clover detector has been studied underground at LUNA and, for comparison, also in an overground laboratory. Spectra have been recorded both for the single segments and for the virtual detector formed by online addition of all four segments. The effect of the escape-suppression shield has been studied as well. 
Despite their generally higher intrinsic background, escape-suppressed detectors are found to be well suited for underground nuclear astrophysics studies. As an example for the advantage of using a composite detector deep underground, the weak ground state branching of the $E_{\rm p}$~=~223\,keV resonance in the $^{24}$Mg(p,$\gamma$)$^{25}$Al reaction is determined with improved precision.
}
\authorrunning{T.\,Sz\"ucs {\it et al.}}

\maketitle

\section{Introduction}

Recent advances in observations \cite[e.g.]{Asplund06-NPA} and in modeling \cite{PenaGaray08-arxiv,Serenelli09-ApJL} of the Sun and of stars have heightened the need for precise nuclear data on reactions of astrophysical interest. One approach to provide such data is to place a high-intensity particle accelerator deep underground, where the laboratory background in $\gamma$-ray detectors is reduced so that radiative capture reactions can be studied with improved sensitivity. 

The Laboratory for Underground Nuclear Astrophysics (LUNA) has implemented this strategy, first with a 50\,kV accelerator \cite{Greife94-NIMA} and now with a 400\,kV accelerator \cite{Formicola03-NIMA} placed in the underground facility of Laboratori Nazionali del Gran Sasso (LNGS)\footnote{Web site of the laboratory: {\tt http://www.lngs.infn.it}} in Assergi, Italy. LNGS is shielded from cosmic rays by a rock overburden equivalent to 3800\,m water. 

Benefiting from the resulting low $\gamma$-ray background, several nuclear reactions of astrophysical importance have been studied in recent years at LUNA \cite{Bonetti99-PRL,Casella02-NPA,Formicola04-PLB,Lemut06-PLB,Bemmerer06-PRL,Confortola07-PRC,Marta08-PRC,Bemmerer09-JPG}. In many cases, cross sections lower than ever reached before have been measured. Motivated by these advances, new underground accelerators have been proposed at a number of locations, namely: LNGS \cite{NuPECC05-Roadmap}, the Canfranc laboratory in Spain\cite{Canfranc09-Workshop}, the planned DUSEL facility in the United States \cite{DOE08-arxiv}, Boulby mine in the United Kingdom \cite{Strieder08-NPA3}, and Romania \cite{Bordeanu08-NPA3}. 

The present work is the third in a series \cite{Bemmerer05-EPJA,Caciolli09-EPJA} that aims to facilitate these efforts, by providing detailed background data on deep underground in-beam setups as a reference case. In the first article of the series, the laboratory background with no or only minor shielding was studied for high-purity germanium (HPGe) and bismuth germanate (BGO) $\gamma$-detectors, and it was shown that for $E_\gamma$~$>$~3\,MeV the laboratory $\gamma$-background at LUNA is typically three orders of magnitude lower than at the surface of the Earth \cite{Bemmerer05-EPJA}. The second article presented an ultra-low background (ULB) HPGe detector with a sophisticated passive shield at LUNA. For $E_\gamma$~$\leq$~3\,MeV, this in-beam setup \cite{Caciolli09-EPJA} displayed a laboratory background close to that of dedicated, deep underground activity-counting setups \cite{Laubenstein04-Apradiso}. 

Here, the effects of segmentation and of active shielding on the laboratory $\gamma$-background of a HPGe detector are studied. To this end, the background of a HPGe detector that has been used for a recent LUNA experiment \cite{Marta08-PRC} has been studied in detail. For some of the experiments, also a 5\,cm thick lead shield has been added, allowing to investigate the combination of active and passive shielding. Finally, as an example of the potential applications of a composite HPGe detector deep underground, the weak branching ratio for the decay of the $E_{\rm p}$~=~223\,keV resonance in the $^{24}$Mg(p,$\gamma$)$^{25}$Al reaction to the ground state in $^{25}$Al is redetermined.

\section{Setup}

\begin{figure}[bt]
\centering
\includegraphics[angle=0,width=1.0\columnwidth]{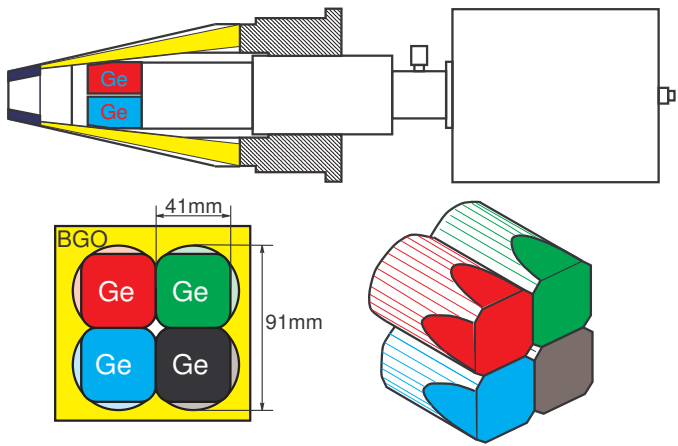}
\caption{Schematic cross section of the Clover-BGO system. The four germanium crystals are called red, green, black, and blue. The BGO escape-suppression shield (yellow) and the heavy-met collimator on the front face (dark-blue) are also shown.}
\label{fig:Clover-scheme}
\end{figure}

For the experiment, a EURISYS Clover detector \cite{Duchene99-NIMA} has been used. 
This type of composite detector was selected because it easily fits in the restricted space of an underground laboratory.
It consists of four coaxial n-type HPGe detectors arranged like a four-leaf clover (fig.~\ref{fig:Clover-scheme}). The spacing between the crystals is only 0.2\,mm, leading to a closely packed geometry. At the 1333\,keV $^{60}$Co line, a single crystal has a typical resolution of 2.2\,keV and 20\% relative efficiency. 

The signals from the four crystals are split after the preamplifiers. One part is fed into four main amplifiers, and the signals are then digitized and recorded in self-triggered, histogramming mode. These four individual histograms were gainmatched and added channel by channel, to form just one histogram hereafter called "singles mode spectrum". 

The second part is fed into an analog summing unit implementing the gain-matching and summing of the four signals. The analog sum signal is then passed to a fifth main amplifier and digitized. The signal can then be recorded either in free-running, self-triggered, mode (called hereafter "addback mode, free-running") or in anticoincidence with the signal from the BGO escape-suppression shield (called hereafter "addback mode, escape \linebreak suppressed"). The virtual large detector formed by the addback mode has 122\% relative efficiency, comparable to the HPGe detectors used for the previous background studies at LUNA \cite{Bemmerer05-EPJA,Caciolli09-EPJA}. 

The accidental suppression rate of the BGO escape-suppression shield was found to be 1\%. The average number of hits per event was determined to be 1.1 for the laboratory background and 1.2 for the highest counting rate in-beam run, on the 278\,keV resonance of the $^{14}$N(p,$\gamma$)$^{15}$O reaction. The timing information from the individual crystals was not used. Due to the continuous character of the intensive ion beam at LUNA, no time correlation between ion beam and emitted $\gamma$-ray was possible.

In the present study, the Clover detector is always used in conjunction with a surrounding BGO scintillator. For the addback mode data, the BGO can act as a Compton suppression veto. For the singles mode data, the BGO was in effect only a passive shield. The detector was used in horizontal geometry, so that in addback mode, the BGO shield can act as a veto against penetrating muons passing the germanium detector volume. 

\section{Off-line experiments and results}

For the underground experiments presented here, the Clover detector was placed deep underground in the LUNA facility \cite{Costantini09-RPP} of LNGS. For a first set of measurements, it was mounted in horizontal geometry at the 45-2 beamline of the 400\,kV LUNA2 accelerator, and no lead shielding was used. During the background measurement presented in the present section, the LUNA beam was off. This setup was used both for the off-line experiments without lead shield and for the in-beam experiment described below in sec.\,\ref{sec:mg24}.

In a second part of the underground experiments, the detector was placed on the floor of the LUNA hall and completely surrounded with a 5\,cm thick shield of standard lead.

For comparison, measurements at the surface of the Earth were performed at FZD. The experimental hall has a ceiling equivalent to about 0.3\,m water. No lead shielding was applied.

\subsection{Laboratory background $\gamma$-lines}

\begin{figure}[t]
\centering
\includegraphics[angle=0,width=1.0\columnwidth]{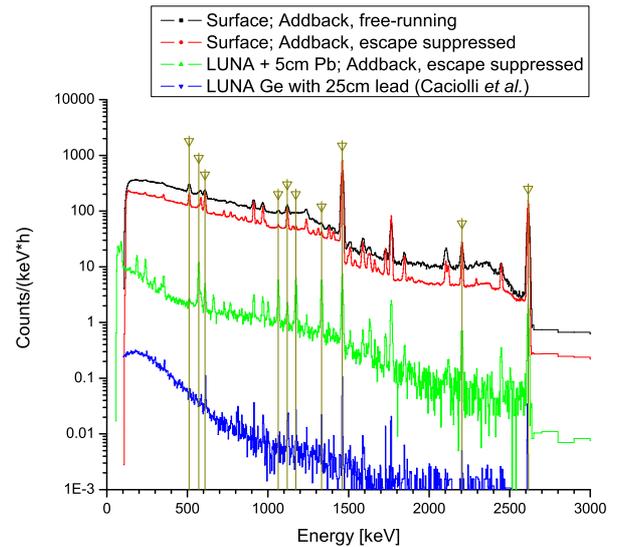}
\caption{Low energy part of the recorded laboratory $\gamma$-background spectra, compared with the previously described, strongly shielded setup at the 45-1 beamline at LUNA \cite{Caciolli09-EPJA}. The lines marked with arrows are discussed in the text.}
\label{fig:spectra_0-3MeV}
\end{figure}

\begin{figure}[t]
\centering
\includegraphics[angle=0,width=1.0\columnwidth]{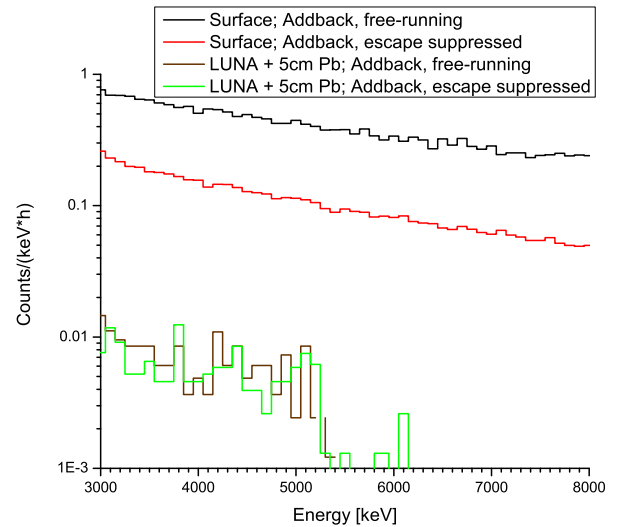}
\caption{High-energy part of the offline $\gamma$-spectra. At LUNA, the escape suppressed and the free-running spectra are undistinguishable in this energy range because the muon flux is so low that the remaining background is not dominated by muons any more, but by neutrons.}
\label{fig:spectra_3-8MeV}
\end{figure}

The main $\gamma$-lines present in the laboratory background (fig.~\ref{fig:spectra_0-3MeV}) are identified as:
\begin{itemize}
\item 511\,keV e$^+$e$^-$ annihilation peak
\item 570 and 1064\,keV lines from $^{207}$Bi. This isotope is a commonly observed contamination in BGO material, produced through the $^{206}$Pb(p,$\gamma$)$^{207}$Bi reaction by cosmic rays. All BGO material shows some $^{207}$Bi impurity, except in cases where the bismuth starting material has been obtained from lead-free ore.
\item 609, 1120, and 2204\,keV lines from the radon daughter $^{214}$Bi. No anti-radon shielding was applied for the present study.
\item 1173 and 1333\,keV from some $^{60}$Co contamination pre\-sent in the BGO crystal.
\item 1461\,keV from $^{40}$K present in the laboratory.
\item 2615\,keV from $^{208}$Tl, in the Thorium chain. The background continuum caused by pileup of the laboratory background reaches up to 5200\,keV, twice the energy of this highest $\gamma$-line (fig.~\ref{fig:spectra_3-8MeV}).
\end{itemize}
Some further lines from radon daughters ($^{228}$Ac and $^{214}$Bi) have also been observed but are neglected in the further discussion because they behave in an analogous manner to the three $^{214}$Bi lines mentioned above.

The counting rates of the above mentioned $\gamma$-lines are summarized in table~\ref{table:gammapeaks}, for the two experiments at LUNA without and with lead shield, and for the reference case at the surface of the Earth. For comparison, the data from the previous study at the 45-1 beamline at LUNA using a single, large HPGe detector with 137\% relative efficiency are also shown \cite{Caciolli09-EPJA}. Those previous data \cite{Caciolli09-EPJA} have been taken in several configurations. Here the previous data taken without shield and those taken with a sophisticated passive shield (25\,cm selected lead with low $^{210}$Pb content, 4\,cm oxygen free high purity copper, anti-radon box) are shown for comparison.

\begin{table*}[tb]
\caption{Measured counting rates for radioactive decay lines, in counts per hour. Present data recorded with the Clover detector, are compared with the previous LUNA values \cite{Caciolli09-EPJA} with a ULB detector and a passive shield consisting of 25\,cm selected lead with low $^{210}$Pb content, 4\,cm oxygen free high purity copper, and an anti-radon box.}
\label{table:gammapeaks}
\resizebox{\textwidth}{!}{
\begin{tabular}{| l | l | r @{ $\pm$ } l | r @{ $\pm$ } l | r @{ $\pm$ } l | r @{ $\pm$ } l | r @{ $\pm$ } l || r @{ $\pm$ } l | r @{ $\pm$ } l | r @{ $\pm$ } l | r @{ $\pm$ } l | r @{ $\pm$ } l |}
	\hline\noalign{\smallskip}
	\multicolumn{2}{|l|}{Source isotope} & \multicolumn{6}{c|}{$^{214}$Bi} & \multicolumn{2}{c|}{$^{40}$K} & \multicolumn{2}{c||}{$^{208}$Tl} & \multicolumn{2}{c|}{e$^+$e$^-$} & \multicolumn{4}{c|}{$^{207}$Bi} & \multicolumn{4}{c|}{$^{60}$Co} \\
	\multicolumn{2}{|l|}{$E_\gamma$ [keV]} & \multicolumn{2}{c|}{609} & \multicolumn{2}{c|}{1120} & \multicolumn{2}{c|}{2204} & \multicolumn{2}{c|}{1461} & \multicolumn{2}{c||}{2615} & \multicolumn{2}{c|}{511} & \multicolumn{2}{c|}{570} & \multicolumn{2}{c|}{1064} & \multicolumn{2}{c|}{1173} & \multicolumn{2}{c|}{1333}\\
	\noalign{\smallskip}\hline\noalign{\smallskip}
	\multirow{3}{*}{\begin{tabular}{@{}l} Clover, \\ Earth's surface \end{tabular}} 
	 & singles & 1013 & 52 & 509 & 26 & 182 & 10 & 6320 & 316 & 1163 & 58 & 1387 & 70 & 239 & 15 & 148 & 10 & 69 & 7 & 61 & 5\\
	 & addback, free runn. & 1405 & 81 & 750 & 47 & 318 & 19 & 10227 & 513 & 2046 & 104 & 1813 & 101 & 387 & 44 & 224 & 30 & 122 & 28 & 78 & 20\\
	 & addback, esc. suppr. & 1415 & 72 & 759 & 39 & 311 & 16 & 10065 & 503 & 1997 & 100 & 800 & 42 & 225 & 17 & 118 & 11 & 84 & 9 & 89 & 8\\
	\noalign{\smallskip}\hline\noalign{\smallskip}
	\multirow{3}{*}{\begin{tabular}{@{}l} Clover at LUNA \\ no shield \end{tabular}} 
	 & singles & 532 & 27 & 258 & 13 & 90 & 5 & 861 & 43 & 284 & 14 & 306 & 16 & 244 & 13 & 149 & 8 & 54 & 3 & 52 & 3\\
	 & addback, free runn. & 750 & 38 & 398 & 20 & 150 & 8 & 1342 & 67 & 481 & 24 & 382 & 21 & 350 & 19 & 225 & 12 & 77 & 6 & 75 & 5\\
	 & addback, esc. suppr. & 717 & 36 & 380 & 19 & 147 & 8 & 1310 & 66 & 459 & 23 & 126 & 8 & 135 & 9 & 59 & 4 & 56 & 4 & 57 & 4\\
	\noalign{\smallskip}\hline\noalign{\smallskip}
	\multirow{3}{*}{\begin{tabular}{@{}l} Clover at LUNA \\ 5\,cm Pb shield \end{tabular}} 
	 & singles & 33 & 3 & 13 & 2 & 5.4 & 0.7 & 42 & 3 & 16.9 & 1.3 & 21 & 3 & 235 & 13 & 153 & 8 & 53 & 3 & 50 & 3\\
	 & addback, free runn. & 51 & 7 & 20 & 4 & 7.9 & 1.5 & 64 & 5 & 28.3 & 2.4 & 34 & 6 & 310 & 18 & 237 & 14 & 86 & 6 & 64 & 5\\
	 & addback, esc. suppr. & 30 & 3 & 15 & 2 & 7.6 & 0.9 & 71 & 4 & 21.0 & 1.6 & 5 & 3 & 98 & 6 & 47 & 3 & 59 & 4 & 58 & 4 \\
	\noalign{\smallskip}\hline\noalign{\smallskip}
	\multirow{2}{*}{ULB at LUNA \cite{Caciolli09-EPJA}} 
	 & no shield & 3729 & 4 & 1278 & 3 & \multicolumn{2}{c|}{} & 4870 & 4 & 1325 & 2 & 762 & 4 & \multicolumn{2}{c|}{} & \multicolumn{2}{c|}{} & \multicolumn{2}{c|}{} & \multicolumn{2}{c|}{} \\
	 & 25\,cm Pb shield & 0.30 & 0.04 & 0.15 & 0.02 & \multicolumn{2}{c|}{} & 0.42 & 0.03 & 0.12 & 0.02 & 0.09 & 0.04 & \multicolumn{2}{c|}{} & \multicolumn{2}{c|}{} & \multicolumn{2}{c|}{} & \multicolumn{2}{c|}{} \\
	\noalign{\smallskip}\hline
  \end{tabular}
}
\end{table*}

As expected, the counting rates of the $\gamma$-lines from radioactive decays are hardly affected by going underground to LUNA because they are dominated by radioisotopes present in the walls of the laboratory or in the detector. When comparing the overground with the unshielded LUNA spectra, it is seen that the radon background ($^{214}$Bi) is a factor two lower at LUNA, due to the better ventilation of the LUNA site. The thorium background ($^{208}$Tl) is lower by a factor four, due to the different characteristics of the rock surrounding the LUNA site, as compared to the FZD hall. A similar effect is observed for the $^{40}$K line and the e$^+$e$^-$ annihilation peak. Only the $\gamma$-lines due to impurities contained in the BGO shield itself ($^{207}$Bi and $^{60}$Co) do not change significantly between the different setups studied, as expected.

When comparing the unshielded and the shielded \linebreak LUNA spectra, it is evident that already the present 5\,cm lead shield leads to sizable reductions in the $\gamma$-line counting rates for all radioisotopes discussed above, except of course for the contaminations inherent to the BGO shield. 

In order to estimate the possible effects of a full, state-of-the-art passive shielding on the present setup, it is useful to compare the present data with the data from the previous LUNA study \cite{Caciolli09-EPJA} (table~\ref{table:gammapeaks}, last two lines). The unshielded starting point of the previous LUNA data is somewhat worse than for the present work, because in the present detector the BGO also acts as passive shield due to its high $\gamma$-attenuation coefficient. However, the factors of improvement seen when comparing the last two lines of table~\ref{table:gammapeaks} show which low levels of background can in principle be reached using a full passive shield like in Ref.~\cite{Caciolli09-EPJA}. 

\subsection{Laboratory background continuum}

\begin{table*}[bt]
\caption{Continuum counting rate in counts/(keV hour) for several regions of interest relevant to radiative capture reactions.}
\label{table:gammacontinuum}
\resizebox{1.0\textwidth}{!}{
\begin{tabular}{| l | l | r @{ $\pm$ } l | r @{ $\pm$ } l | r @{ $\pm$ } l| r @{ $\pm$ } l| r @{ $\pm$ } l|| r @{ $\pm$ } l |}
	\hline\noalign{\smallskip}
	\multicolumn{2}{|l|}{Reaction} & \multicolumn{2}{c|}{ $^{12}$C($^{12}$C,p)$^{23}$Na} & \multicolumn{2}{c|}{$^{2}$H($\alpha$,$\gamma$)$^{6}$Li} & \multicolumn{2}{c|}{$^{3}$He($\alpha$,$\gamma$)$^{7}$Be} & \multicolumn{2}{c|}{$^{12}$C(p,$\gamma$)$^{13}$N} & \multicolumn{2}{c||}{$^{24}$Mg(p,$\gamma$)$^{25}$Al} & \multicolumn{2}{c|}{$^{14}$N(p,$\gamma$)$^{15}$O}\\
	\multicolumn{2}{|l|}{$\gamma$-ray ROI [keV]} & \multicolumn{2}{c|}{425-455} & \multicolumn{2}{c|}{1545-1575} & \multicolumn{2}{c|}{1738-1753} & \multicolumn{2}{c|}{2004-2034} & \multicolumn{2}{c||}{2470-2500} & \multicolumn{2}{c|}{6000-8000} \\
	\noalign{\smallskip}\hline\noalign{\smallskip}
	\multirow{3}{*}{\begin{tabular}{@{}l} Clover, \\ Earth's surface \end{tabular}}
	& singles & 259.4 & 0.2 & 16.02 & 0.06 & 10.79 & 0.07 & 8.48 & 0.04 & 3.81 & 0.03 & (128.9 & 0.8)$\times 10^{-3}$ \\
	& addback, free running & 317.3 & 0.8 & 22.64 & 0.21 & 15.84 & 0.25 &  12.67 & 0.16 & 6.33 & 0.05 & (205.6 & 1.1)$\times 10^{-3}$\\
	& addback, escape suppressed & 162.2 & 0.2 & 11.48 & 0.05 & 8.36 & 0.06 &  5.75 & 0.04 & 4.07 & 0.03 & (19.2 & 0.4)$\times 10^{-3}$ \\
	\noalign{\smallskip}\hline\noalign{\smallskip}
	\multirow{3}{*}{\begin{tabular}{@{}l} Clover at LUNA \\ no shield \end{tabular}}
	& singles & 64.57 & 0.14 & 5.29 & 0.04 & 3.09 & 0.04 &  2.17 & 0.02 & 0.72 & 0.01 & (0.17 & 0.03)$\times 10^{-3}$ \\
	& addback, free running & 77.64 & 0.18 & 7.52 & 0.06& 4.98 & 0.07 &  3.13 & 0.04 & 1.27 & 0.02 & (0.15 & 0.04)$\times 10^{-3}$ \\
	& addback, escape suppressed & 30.52 & 0.14 & 3.12 & 0.04 & 2.71 & 0.06 &  1.17 & 0.03 & 0.73 & 0.02 & (0.18 & 0.05)$\times 10^{-3}$ \\
	\noalign{\smallskip}\hline\noalign{\smallskip}
	\multirow{3}{*}{\begin{tabular}{@{}l} Clover at LUNA \\ 5\,cm Pb shield \end{tabular}} 
	& singles & 6.35 & 0.11 & 0.40 & 0.03 & 0.19 & 0.03 &  0.130 & 0.015 & 0.06 & 0.01 & (0.28 & 0.12)$\times 10^{-3}$ \\
	& addback, free running & 6.95 & 0.17 & 0.57 & 0.05 & 0.34 & 0.05 &  0.21 & 0.03 & 0.11 & 0.02 & \multicolumn{2}{c|}{$<$ 0.11$\times 10^{-3}$} \\
	& addback, escape suppressed & 2.47 & 0.07 & 0.26 & 0.02 & 0.16 & 0.03 &  0.076 & 0.013 & 0.05 & 0.01 & (0.29 & 0.13)$\times 10^{-3}$ \\
	\noalign{\smallskip}\hline\noalign{\smallskip}
	HPGe at LUNA & 5\,cm Pb shield \cite{Bemmerer05-EPJA} & \multicolumn{2}{c|}{} & \multicolumn{2}{c|}{} & \multicolumn{2}{c|}{} & \multicolumn{2}{c|}{} & \multicolumn{2}{c||}{} & \multicolumn{2}{c|}{$<$ 0.1$\times 10^{-3}$} \\
	ULB at LUNA & 25\,cm Pb shield \cite{Caciolli09-EPJA} & 0.072 & 0.002 & 0.0015 & 0.0003 & 0.0011 & 0.003 & 0.0009 & 0.0003 & \multicolumn{2}{c||}{} & \multicolumn{2}{c|}{} \\
	\noalign{\smallskip}\hline
  \end{tabular}
}
\end{table*}

For in-beam experiments, the $\gamma$-ray continuum observed in regions outside of the laboratory background lines is of paramount importance.
For reaction $Q$-values above 3\,MeV, in principle also $\gamma$-rays of energies above 3\,MeV can be emitted, in a region where there are no $\gamma$-lines from radioisotopes. Furthermore, for primary in-beam $\gamma$-rays the resolution is in many cases not limited by the detector, but by the effective target thickness, making the $\gamma$-lines rather wide, adding further importance to obtaining a low continuum in $\gamma$-detectors.

At the surface of the Earth, the two main sources of the $\gamma$-continuum are the Compton continuum of $\gamma$-rays and the energy loss or stopping of cosmic-ray induced particles like muons. An escape-suppression veto detector like the present BGO shield can strongly reduce both effects. In addition, placing the setup deep underground, thus reducing the muon flux, should lead to a further reduction of the $\gamma$-continuum, both for $E_\gamma$~$<$~3\,MeV \cite{Caciolli09-EPJA} and $E_\gamma$~$\geq$~3\,MeV \cite{Bemmerer05-EPJA}. For example, a previous Monte Carlo simulation \cite{Vojtyla06-RITE} predicts an overall factor of three reduction for  $E_\gamma$~$<$~3\,MeV, when comparing overground spectra with a shallow underground facility at a depth of 30\,m water equivalent.

In order to verify these expectations, the continuum counting rate has been determined for some regions of interest (ROI's) that are important for nuclear reactions that might conceivably be studied in underground accelerator experiments (table~\ref{table:gammacontinuum}). These reactions and the astrophysical motivation driving their study have been discussed previously \cite{Caciolli09-EPJA}. 

For $E_\gamma$~$<$~3\,MeV, it is clear from table~\ref{table:gammacontinuum} that the present detector, which has some internal contamination and is at maximum shielded with 5\,cm lead, cannot reach the background suppression factors of the previous LUNA study \cite{Caciolli09-EPJA} with its much better shield (table~\ref{table:gammacontinuum}, last line). 

For $E_\gamma$~$\geq$~3\,MeV overground, it is found from the present data that the escape suppression reduces the continuum counting rate by a factor 11.
This reduction is comparable to the factor 10--50 reported for $E_\gamma$ = 7--11\,MeV from a previous overground experiment using a HPGe detector shielded by a NaI escape-suppression shield \cite{Mueller90-NIMA}. 

By placing the detector deep underground at LUNA, in the same energy region the continuum counting rate is improved by an additional factor of 30 when compared with the overground, escape suppressed run (fig.~\ref{fig:spectra_3-8MeV}). For 2.6\,MeV~$<$~$E_\gamma$~$<$~5.2\,MeV (two times the energy of the $^{208}$Tl $\gamma$-ray), the LUNA spectra are dominated by pileup from natural radionuclides. This background is not affected by the BGO veto detector, but it can instead be rejected using suitable electronic pileup rejection logic. However, for LUNA-type experiments such circuits may lead to increased uncertainty, because at low counting rate it is not easy to properly adjust them. Therefore, no pileup rejection circuit is used here.

At LUNA, the escape suppression does not produce any further effect for $E_\gamma$~$>$~5.2\,MeV, as expected when muons make a negligible contribution to the background (table~\ref{table:gammacontinuum}). Similarly, the 5\,cm lead shield does not lead to a further reduction in counting rate at LUNA, which can be explained by the fact that radioisotopes don't contribute significantly to the background for $E_\gamma$~$\geq$~5.2\,MeV. The remaining background values shown for the present detector are consistent with the previous data for a similar germanium detector with 5\,cm lead shield at LUNA \cite{Bemmerer05-EPJA}. This background level is explained with neutron capture from the remaining flux of thermal and high-energetic neutrons present in LNGS \cite{Belli89-NCA}.

\subsection{Addback factor}

For the present data on the laboratory background lines (table~\ref{table:gammapeaks}), the addback factor \cite{Duchene99-NIMA}
\begin{equation}
{\rm ABF} \stackrel{!}{=} \frac{C_{\rm addback, free\!-\!running}}{C_{\rm singles}}
\label{eq:addbackfactor}
\end{equation}
has been calculated. Here, $C_{\rm addback, free\!-\!running}$ is the counting rate in addback mode, free-running, and $C_{\rm singles}$ is the singles mode counting rate. The same has been done also for some $\gamma$-lines emitted in the $^{14}$N(p,$\gamma$)$^{15}$O reaction studied with the present detector and setup. 

\begin{figure}
\centering
\includegraphics[angle=0,width=1.0\columnwidth]{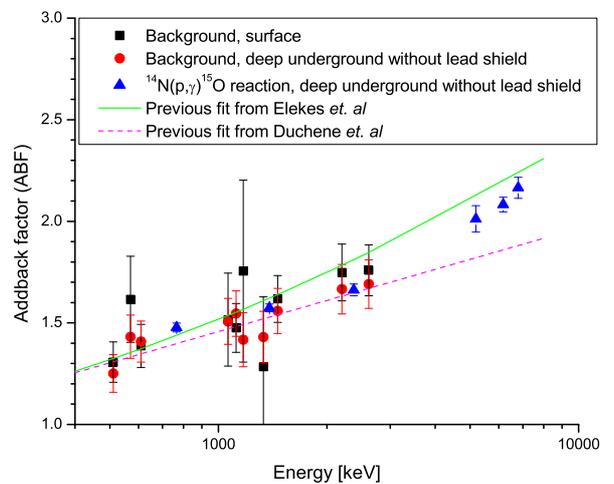}
\caption{Symbols, addback factor ABF calculated according to eq.~\ref{eq:addbackfactor} for $\gamma$-lines from the laboratory background (table~\ref{table:gammapeaks}): Squares, Earth's surface; circles, deep underground without lead shield. Triangles, ABF for $\gamma$-lines from the the $^{14}$N(p,$\gamma$)$^{15}$O reaction. Solid (dashed) line, previous fitted curves from Ref.~\cite{Elekes03-NIMA} (from Ref.~\cite{Duchene99-NIMA}).}
\label{fig:addbackfactor}
\end{figure}

The data points all follow the same general curve, despite the very different points of emission of the various $\gamma$-rays: outside contaminations, radioactivity in the BGO shield, or decays in the air close to the detector (fig.~\ref{fig:addbackfactor}). The present high-energy data points lie close to the previous fitted curve \cite{Elekes03-NIMA}, confirming that the slope is somewhat higher than initially expected \cite{Duchene99-NIMA}.

\section{Decay of the $E_{\rm p}$~=~223\,keV resonance in the $^{24}$Mg(p,$\gamma$)$^{25}$Al reaction as an example}
\label{sec:mg24}

\subsection{General considerations}

\begin{figure}
\centering
\includegraphics[angle=0,width=0.8\columnwidth,height=0.3\textheight,keepaspectratio,trim=10mm 5mm 2mm 23mm,clip]{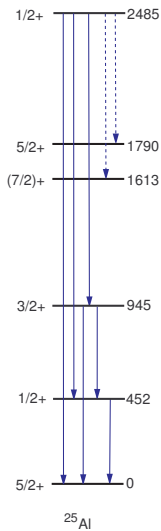}
\caption{Level scheme of $^{25}$Al \cite{Firestone09-NDS25}. Full arrows, $\gamma$-transitions observed in the present experiment. Dashed arrows, $\gamma$-transitions where new upper limits have been derived.}
\label{fig:25Al}
\end{figure}

When studying the $\gamma$-decay of an excited nuclear state (e.g. the $E_{\rm x}$~=~2485\,keV state in $^{25}$Al, fig.~\ref{fig:25Al}), usually not only direct decay to the ground state of the nucleus, but also cascade decays via intermediate states are observed. Therefore in the observed $\gamma$-ray spectrum the signal from the transition to the ground state can be obscured by an artefact that appears at exactly the same energy, due to the true coincidence summing effect. This effect is usually corrected for in an analytic manner. However, in cases where the summing-in effect is large when compared to the true signal, such a correction can lead to considerable systematic uncertainty. 

The magnitude of the summing-in correction is directly proportional to the absolute $\gamma$-detection efficiency. Therefore, one possible approach to limit summing-in is to move the detector to a larger distance, sacrificing efficiency and angular coverage. However, in low-energy nuclear astrophysics experiments, usually the $\gamma$-ray emission rate is low and their angular distribution not very well known. Solving the summing problem in this way therefore worsens two other problems, the low statistics and the dependence on the angular distribution. Therefore in the past at LUNA this approach could only be used for data on strong resonances \cite{Imbriani05-EPJA}.

An alternative approach is to use a composite detector. For the present case of four independent crystals, the summing-in effect is reduced by 4$\cdot$${\rm ABF}$ (i.e. four times the addback factor, ${\rm ABF}$), while the $\gamma$-efficiency is only reduced by ${\rm ABF}$. The angular coverage even remains unchanged. As an additionial piece of information, the addback data can also be analyzed, and the comparison of singles and addback mode data can serve as a check on the analytical summing correction for the addback data.

A further advantage of using a composite detector, the much lower Doppler correction for each single crystal, has only limited importance for low-energy nuclear astrophysics studies. For Gamow peak energies of a few ten keV, the typical velocity of the recoil nuclei is lower than 1\%, and the typical Doppler correction for LUNA-type experiments is of the same order as the energy resolution of the HPGe detector.

\subsection{Branching ratio determination}

In order to illustrate these considerations, the weak ($\approx$3\%) ground state branching of the $E_{\rm p}$~=~223\,keV resonance in the $^{24}$Mg(p,$\gamma$)$^{25}$Al reaction (corresponding to the $E_{\rm x}$ = 2485.3\,keV level in $^{25}$Al, fig.~\ref{fig:25Al}) is redetermined here. This reaction plays a role in the hydrogen-burning MgAl chain \cite{Iliadis07-Book}. 

\begin{figure*}
\centering
\includegraphics[angle=0,width=1.0\textwidth,trim=60mm 10mm 65mm 10mm,clip]{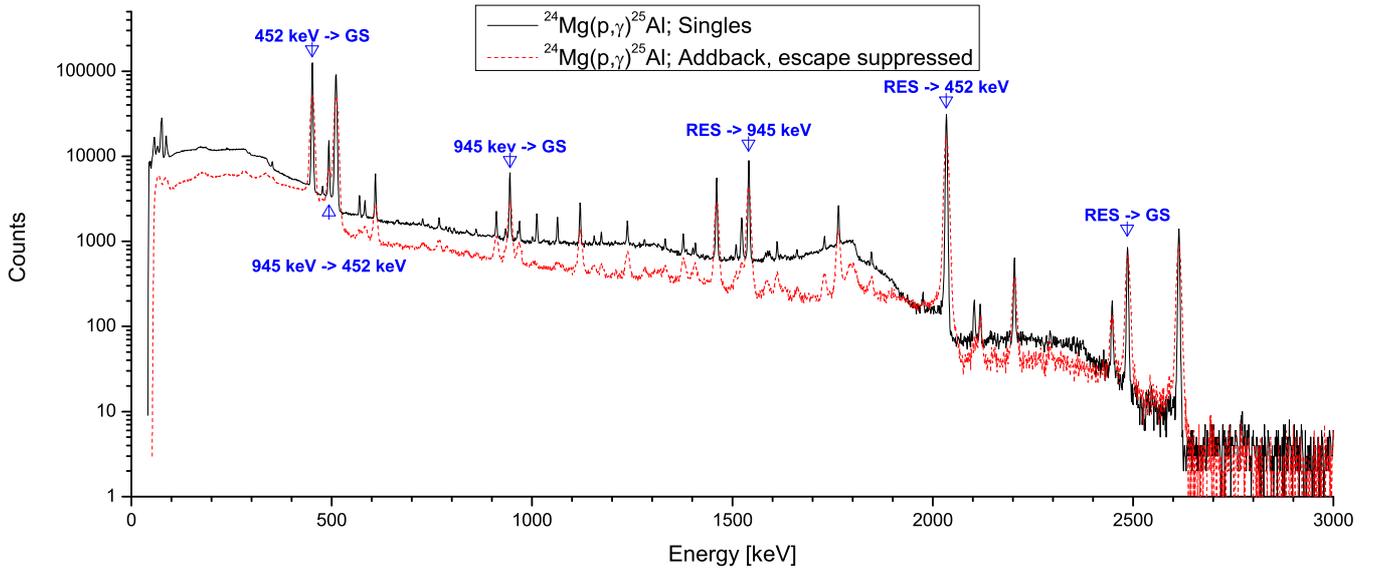}
\caption{In-beam $\gamma$-spectrum on the top of the $E_{\rm p}$~=~223\,keV resonance in the $^{24}$Mg(p,$\gamma$)$^{25}$Al. Black full (red dashed) line, singles mode (addback mode, escape suppressed) data. The most important transitions are marked.}
\label{fig:Mg}
\end{figure*}

For the experiment, a magnesium oxide target of natural isotopic composition (79\% $^{24}$Mg) was used. The Clover detector was placed at 55$^\circ$ with respect to the ion beam, with its front face at 9.5\,cm distance from the target. The $\gamma$-detection efficiency is well-known from another experiment in exactly the same geometry \cite{Marta08-PRC}, and the slope from 695\,keV to 2485\,keV is known to 1.0\%. By scanning the target profile, an energy near the center of the target was selected. Then a spectrum was recorded on top of the resonance (fig.~\ref{fig:Mg}). With a strength of (12.7$\pm$0.9)\,meV \cite{Powell99-NPA}, the resonance is sufficiently intensive that off-resonance capture can be neglected for the present purposes. The laboratory background is comparable in intensity to the in-beam lines, as is apparent from the similar yield of the in-beam line at 2485\,keV and the laboratory background line at 2615\,keV (fig.~\ref{fig:Mg}). However, the background $\gamma$-lines lie at different energies, so the background does not limit the statistics of the 2485\,keV ground state line (table~\ref{table:gammacontinuum}).

The branching ratios for the decay of the resonance have then been determined (table~\ref{table:mg24branching}). For the ground state capture line, the calculated summing-in correction was 37\% (7\%) for the addback (singles) mode data, respectively. Assuming a conservative 20\% relative uncertainty for all the summing-in and summing-out corrections, due to the summing correction there is 0.19\% (0.04\%) absolute uncertainty in the ground state branching for addback (singles) mode. For the addback case, this dominates the total uncertainty of 0.2\%. The fact that the branching ratio as determined in the addback mode agrees with the singles mode data confirms that the summing-in correction is accurate. 

For the primary $\gamma$-ray from the major transition, capture to the 452\,keV first excited state, 3\% (0.7\%) summing-out correction was taken into account for addback (singles) mode. For the primary $\gamma$-ray from capture to the 945\,keV state, 5\% (1.0\%) summing-out correction was taken into account, and again the addback and singles data are in agreement. 

The newly determined branching ratios are in agreement with the literature data \cite{Powell99-NPA} but more precise. No significant branching is expected for the M3 transition to the $\frac{7}{2}^+$ level at 1613\,keV and the E2 transition to the $\frac{5}{2}^+$ level at 1790\,keV. The present data bear out this expectation, giving new experimental upper limits for these two transitions (table~\ref{table:mg24branching}). The values obtained in singles mode are recommended for future compilations \cite{Firestone09-NDS25}.

In the previous measurement \cite{Powell99-NPA}, a large volume \linebreak (140\%) HPGe detector had been placed at 55$^\circ$ with respect to the beam direction, at 5.9\,cm distance to the target. 
Based on these numbers, we estimate that in singles mode, the present summing-in correction is about a factor 9 lower than in Ref.~\cite{Powell99-NPA}, justifying the present lower uncertainty.  



\begin{table}[t]
\caption{Branching ratios, in \%, for the decay of the $E_{\rm p}$~=~223\,keV resonance in the $^{24}$Mg(p,$\gamma$)$^{25}$Al reaction. Where applicable, upper limits are given for 90\% confidence level.}
\label{table:mg24branching}
\resizebox{1.0\columnwidth}{!}{
\begin{tabular}{|r@{ $\rightarrow$ }l|*{3}{r@{$\pm$}l|}}
\multicolumn{2}{|c|}{Decay} & \multicolumn{2}{c|}{Literature} & \multicolumn{4}{c|}{Present work} \\ 
\multicolumn{2}{|c|}{} & \multicolumn{2}{c|}{\cite{Powell99-NPA}} & \multicolumn{2}{c|}{addback} & \multicolumn{2}{c|}{singles} \\ \hline
2485, $\frac{1}{2}^+$&0, $\frac{5}{2}^+$ & 2.7&0.3 & 2.6&0.2 & 2.69&0.08 \\
&452, $\frac{1}{2}^+$ & 81.7&3.4 & 81.8&1.2 & 81.6&1.1\\
&945, $\frac{3}{2}^+$ & 15.6&1.1 & 15.6&0.5 & 15.7&0.6\\
 &1613, $\frac{7}{2}^+$ & \multicolumn{2}{c|}{$<$0.8} & \multicolumn{2}{c|}{$<$0.3} & \multicolumn{2}{c|}{$<$0.3} \\
 &1790, $\frac{5}{2}^+$ &  \multicolumn{2}{c|}{$<$0.8} & \multicolumn{2}{c|}{$<$0.3} & \multicolumn{2}{c|}{$<$0.3} \\[1mm] \hline
\end{tabular}
}
\end{table}

Another example studied recently is the $^{14}$N(p,$\gamma$)$^{15}$O reaction, which controls the rate of the hydrogen-burning CNO cycle \cite{Iliadis07-Book}. Due to the complicated interference pattern of several components in the R-matrix framework, the rather weak capture to the ground state in $^{15}$O dominates the uncertainty of the total extrapolated $^{14}$N(p,$\gamma$)$^{15}$O cross section at energies corresponding to solar hydrogen burning. The study of this transition is affected by summing-in corrections, and with the present detector and setup recently an experiment with greatly reduced summing corrections has been performed \cite{Marta08-PRC}.

\section{Discussion and outlook}

A Clover-BGO detector system for nuclear astrophysics experiments has been used deep underground at LUNA. The laboratory background of one and the same detector has been studied in detail at LUNA and in an overground laboratory for reference. It is found that by going deep underground, the $\gamma$-continuum background counting rate can be reduced much more than by simply applying a cosmic-ray veto. 

In free-running mode, the background characteristics of the present detector at LUNA are comparable to single detectors of similar size at LUNA, when a shielding similar to the present one is applied. The escape suppression was shown to further reduce the $\gamma$-continuum background counting rate.  

In order to illustrate the applications of a composite, escape-suppressed detector in underground nuclear astrophysics, the weak ground state branching of the \linebreak $E_{\rm p}$~=~223\,keV resonance in the $^{24}$Mg(p,$\gamma$)$^{25}$Al reaction has been determined with improved precision. 

A further step in studying the potential of a composite, escape-suppressed detector in a deep underground accelerator laboratory such as LUNA would be to construct an ultra-low background composite detector with a long neck to accommodate a full lead and copper shield.

\section*{Acknowledgments}

The present work has been supported by INFN and in part by the EU (ILIAS-TA RII3-CT-2004-506222), OTKA (T49245 and K68801), and DFG (Ro~429/41). T.S. acknowledges a Herbert Quandt fellowship at Technical University Dresden.


\end{document}